\documentclass{elsart4-1}
\usepackage{amssymb}
\usepackage[english,francais]{babel}
\usepackage{graphicx}

\newcommand{\eps} {\varepsilon}

\begin{document}
% Select a primary header Physics or Astrophysics
% You can place after the header (classification), if you know it.

\centerline{Physics Header}
\begin{frontmatter}

\selectlanguage{english}
\title{Phase-separation of miscible liquids in a centrifuge}

% If all authors are at the same address, the [label1] can be suppressed

\selectlanguage{english}
\author[authorlabel1]{Yoav Tsori},
\ead{tsori@bgu.ac.il}
\author[authorlabel2]{Ludwik Leibler}
\ead{ludwik.leibler@espci.fr}

\address[authorlabel1]{Department of Chemical Engineering, Ben-Gurion University of the
Negev, \\
P.O. Box 653, 84105 Beer-Sheva, Israel}
\address[authorlabel2]{Laboratoire Mati\`ere Molle \& Chimie (UMR 7167), ESPCI,\\
10, rue Vauquelin, 75231 Paris CEDEX 05, France}

\medskip
\begin{center}
{\small Received *****; accepted after revision +++++}
\end{center}

\begin{abstract}
We show that a liquid mixture in the thermodynamically stable homogeneous phase can 
undergo a
phase-separation transition when rotated at sufficiently high frequency $\omega$. This
phase-transition is different from the usual case where two liquids are immiscible or
where the slow sedimentation process of one component (e.g. a polymer) is accelerated due
to centrifugation. For a binary mixture, the main coupling is due to a term $\propto
\Delta\rho(\omega r)^2$, where $\Delta\rho$ is the difference between the two liquid
densities and $r$ the distance from the rotation axis. Below the critical temperature
there is a critical rotation
frequency $\omega_c$, below which smooth density gradients occur. When 
$\omega>\omega_c$,
we find a sharp interface between the low density liquid close to the center of the
centrifuge and a high density liquid far from the center. These findings may be relevant
to various separation processes and to the control of chemical reactions, in particular
their kinetics.
\vskip 0.5\baselineskip

\selectlanguage{francais}
\vskip 0.5\baselineskip
\noindent
{\bf S\'{e}paration de phases des liquides miscibles dans une centrifugeuse }\\
\noindent{\bf R\'esum\'e}\\
\noindent
%Your r\'esum\'e in French here.
Nous prouvons qu'un m\'{e}lange de liquides dans une phase homog\`{e}ne
thermodynamiquement stable peut subir une transition de phase une fois
centrifug\'{e} \'{a} une fr\'{e}quence $\omega$ suffisamment \'{e}lev\'{e}e.
Cette transition de phase est diff\'{e}rente du cas habituel o\`{u} deux liquides sont
non-miscibles ou du cas o\`{u} un processus lent de s\'{e}dimentation d'un composant (par 
exemple un polym\`{e}re) est acc\'{e}l\'{e}r\'{e} par une centrifugation. Pour un m\'{e}lange 
binaire, le couplage principal est d\^{u} \`{a} un terme $\propto\Delta\rho(\omega r)^2$, 
o\`{u} $\Delta\rho$ est la diff\'{e}rence des densit\'{e}s entre les deux liquides
et $r$ est la distance de l'axe de rotation. Au-dessous de la temp\'{e}rature critique, 
il y a une rotation critique de
fr\'{e}quence $\omega_c$; au-dessous d'$\omega_c$, de faibles gradients de densit\'{e} 
prennent naissance.
Quand $\omega>\omega_c$,
nous trouvons une interface mince entre le liquide de faible densit\'{e} pr\`{e}s du
centre de la centrifugeuse et un liquide \`{a} haute densit\'{e} loin du centre. Ces
r\'{e}sultats peuvent \^{e}tre appropri\'{e}s
\`{a} divers processus de s\'{e}paration et au contr\^ole de
r\'{e}actions chimiques, en particulier de leur cin\'{e}tique.

%Now keywords/mots-cl�
\vskip 0.5\baselineskip
\noindent{\small{\it Mots-cl\'es~:} Centrifugeuse; transition des phases; 
fr\'{e}quence critique}
}
\end{abstract}
\end{frontmatter}

% % now the Version fran�ise abr��, if it exists
% \selectlanguage{francais}
% \section*{Version fran\c{c}aise abr\'eg\'ee}
% % Text of your Version fran�ise abr�� here

\selectlanguage{english}
% main text
\section{Introduction}
% \label{}
% etc, etc
The phase of a liquid mixture typically depends on the ambient
temperature, pressure, and composition, but may also be influenced by
external fields such as magnetic and electric fields
\cite{LL_stat,onuki1,onuki2,TTL,TL_pnas}, surface fields \cite{TA_int_sci,YT_inter_dig}, 
or gravity \cite{khokhlov,chaikin1,chaikin2,TL_pre2005,sengers1,leung}. 
The gravitational field couples to the density
of the different components, and thus tends to separate them 
\cite{greer81,greer75,moldover79}. Since this coupling is rather
weak, gravity is mainly associated with the slow sedimentation process of an already
immiscible mixture. A common way to accelerate this sedimentation is by the use of
centrifugation, where the high achievable rotation frequencies permit an
effective
acceleration highly superior to the simple gravitation case \cite{boyum,esterman79}. 
Since centrifugation is very
effective as an aid to separation of substances which are not thermodynamically 
mixed, it
is rarely used on thermodynamically homogeneous mixtures. Indeed, as we show below, 
when a
homogeneous mixture is rotating around some symmetry axis, only small 
composition
gradients appear in the radial direction with respect to this axis \cite{esterman79}. 
However, and this is a
surprising result of our analysis, if the mixture is rotating rapidly enough, above a
threshold value, the mixture composition exhibits a sharp jump from one composition to
another; this sharp change is the signature of a phase-separation. This phase separation
should be accessible for many mixtures using modern ultracentrifugation
techniques. A simple theory is 
developed, analyzed and discussed below.

\section{Theory and Results}
The mixture under consideration here is binary and symmetrical, consisting of two polymers
A and B, each made up of $N$ identical monomers of volume $v_0$. The case of simple
liquids is obtained when $N=1$. The local volume fraction of the A
molecules is $\phi$ ($0<\phi<1$), and the dimensionless Flory parameter
$\chi\sim 1/T$ characterizes the repulsion between the species. 
The average mixture composition is denoted $\phi_0$ and the mass density is $\rho_0$. We
further assume a linear constitutive relation between composition and density, namely
\begin{eqnarray}
\rho(\phi)=\rho_0+\Delta\rho(\phi-\phi_0)
\end{eqnarray} 
where $\Delta\rho\equiv \rho_A-\rho_B$, and $\rho_A$ and $\rho_B$ are the densities of
the A and B components, respectively. 
\footnote{Our theory does not depend on the linear relation between mass density
and composition. Quadratic and higher order dependences lead to qualitatively similar
conclusions.}
The starting point of our formulation for a mixture rotating in an angular
frequency $\omega$ around some axes is the following free-energy density 
\begin{eqnarray}\label{FE}
f=\frac{k_BT}{v_0}
f_m(\phi)-\frac12\rho(\phi)(\omega r)^2
\end{eqnarray} 
In the above $f_m$ is the Flory-Huggins free energy of mixing,
\begin{eqnarray}\label{fm}
f_m(\phi)=
\frac1N\phi\ln(\phi)+\frac1N(1-\phi)\ln(1-\phi)+\chi\phi(1-\phi)+\frac12\xi^2(\nabla\phi)^
2
\end{eqnarray} 
where $k_B$ is the
Boltzmann constant and $T$ is the temperature. The last term, proportional to $(\nabla
\phi)^2$, is due to an energy penalty for having composition gradients. It gives
rise to a characteristic length scale of phase-separated domains scaling as
$\sim(T_c-T)^{-1/2}$, diverging at the critical point \cite{safran}. In the discussion
below, we are not restricted to mixtures close to the critical point, and therefore this
term is subsequently neglected.
The phase-diagram derived from this
mixing energy has a critical point at $(\phi_c,\chi_c)=(1/2,2/N)$. The transition
(binodal)
value of $\chi$ at a given composition $\phi$ is given by
\begin{eqnarray}
N\chi_t=\frac{1}{2\phi-1}\ln\left(\frac{\phi}{1-\phi}\right)
\end{eqnarray} 
We may write to a good
approximation $T=B/\chi$, where $B$ is a constant with units of K, and in this case the
critical temperature is $T_c=BN/2$ and the transition temperature is $T_t=B/\chi_t$.
We are interested in mixtures in the homogeneous phase, at temperatures $T$ below $T_c$
but above $T_t$, and off-critical compositions ($\phi\neq 1/2$).
The second term in Eq. (\ref{FE}), proportional to $(\omega
r)^2$, is due to the centrifugal forces. Here $r$ is the distance from the rotation
axes, and is bounded by the smallest and largest radii: $R_1<r<R_2$.
\begin{figure}[h!]
\begin{center}
\includegraphics[scale=0.5,bb=35 180 550 585,clip]{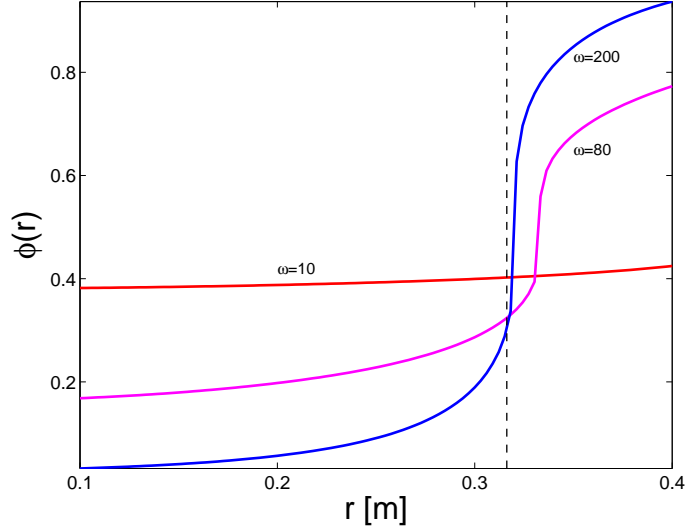}
\end{center}
\caption{Composition profiles for a binary liquid mixture confined in a
tube with inner
radius $R_1=10$ cm and outer radius $R_2=40$ cm, as obtained from Eqs. 
(\ref{gov_eqn1}) and (\ref{gov_eqn2}) with $\chi_c<\chi<\chi_t$ ($T_c>T>T_t$).
At low rotation frequencies
$\omega=10$ s$^{-1}$, the high-density component is pulled toward the larger radius, and
the
resulting profile exhibits weak gradients. If the rotation frequency is large enough,
$\omega=80$ s$^{-1}>\omega_c$, a sharp interface separating large and small
$\phi$-values
appears. As the frequency is further increased to $\omega=300$ s$^{-1}$, the difference
between
coexisting domains becomes larger, and the front moves to smaller
values of $r$. The horizontal dashed line is the lowest possible value of the front
location $R_{\rm min}$ [Eq. \ref{Rmin})], 
corresponding to maximal phase-separation. We used $N=200$,
$v_0=2\times
10^{-28}$ m$^3$, $\rho_0=1000$ Kg/m$^3$, $\Delta\rho=100$ Kg/m$^3$, $\phi_0=0.4$,
$T_c=80^\circ$C and the temperature is $T-T_t=3.5^\circ$K.
}
\end{figure}

In the presence of Earth's gravity alone, such an inertia term is absent and one has
to consider buoyancy effects only very close to
the consolute point.  However, as we will show below and in particular for modern
centrifuges rotating at high speeds, these inertia effects are important and can modify
the thermodynamic aspects of separation phenomena.

We minimize the free-energy in Eq. (\ref{FE}) to obtain the following governing
equation for the profile $\phi({\bf r})$:
\begin{eqnarray}\label{gov_eqn1}
\frac{k_BT}{Nv_0}\left[
\ln\left(\frac{\phi}{1-\phi}\right)+N\chi(1-2\phi)\right] -\frac12\Delta\rho(\omega
r)^2-\mu=0
\end{eqnarray} 
$\mu$ is the Lagrange multiplier (chemical potential) required to conserve the average
composition of a system enclosed in a volume $V$:
\begin{eqnarray}\label{gov_eqn2}
\frac1V\int \phi({\bf r}){\rm d}^3r=\phi_0
\end{eqnarray}
Eq. (\ref{gov_eqn1}) gives an analytical inverted expression for the profile $r$ as a
function of $\phi$.
\begin{figure}[h!]
\begin{center}
\includegraphics[scale=0.5,bb=30 180 560 580,clip]{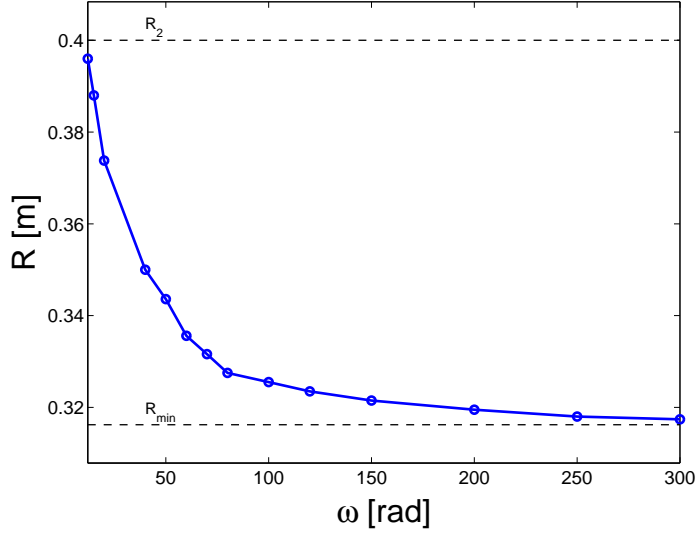}
\end{center}
\caption{Equilibrium front location $R$ separating high- and low-$\phi$
domains as a
function of angular frequency $\omega$, as obtained from Eqs. (\ref{gov_eqn1}) and
(\ref{gov_eqn2}).
$R$ is smaller than $R_2$ and decreases as
$\omega$ increases above $\omega_c\simeq 12$ s$^{-1}$; it tends to $R_{\rm min}$ [Eq.
(\ref{Rmin})] for very
large $\omega$. The approximate expression for $\omega_c$, Eq. (\ref{wc}), gives 
$\omega_c=15$ s$^{-1}$.
All parameters and constants are the same as in Fig. 1.
}
\end{figure}

In Fig. 1 we display the calculated profile $\phi(r)$ for several angular frequencies
$\omega$.
Before centrifugation, the mixture is initially homogeneous, and thus $\chi<\chi_t$
($T>T_t$).
The centrifugal term in Eq. (\ref{gov_eqn1}) is
spatially-dependent, and this means that when the mixture is rotated, $\phi(r)$ has
smooth density changes: high-density material is pulled toward large radii whereas 
small
density material is pulled toward small $r$. This behavior is found for $\chi<\chi_c$,
irrespective of $\omega$. A different scenario occurs for $\chi_c<\chi<\chi_t$, which
corresponds to a temperature lower than the critical temperature but higher than the
binodal. At this regime, if $\omega$ is small, smooth composition variations occur.
However, there exist a critical rotation frequency $\omega_c$,
above which a phase-separation transition takes place. If $\omega>\omega_c$, 
$\phi(r)$ has a 
jump from large values at large $r$ to low values at small $r$. Note that
$\phi(r)$ is not constant in each of the domains, but instead has a weak gradient, as 
is seen in Fig. 1. The equilibrium
location of the front separating the two domains, $R$, moves to smaller $r$ values as 
the
rotation frequency is further increased above $\omega_c$. As $R$ decreases, the
difference between the compositions in the two domains increases. $R$ has a lower bound
denoted $R_{\rm min}$, which corresponds to the maximal possible separation between the
two liquids: pure A component is found at large $r$'s and pure B is closer to the
centrifuge's axes. It follows that $R_{\rm min}$ obeys
\begin{eqnarray}\label{Rmin}
R_{\rm min}^2=R_2^2(1-\phi_0)+R_1^2\phi_0
\end{eqnarray} 

In order to obtain an analytical expression for the critical
frequency $\omega_c$, we continue with the following approximation. We expand Eq. 
(\ref{gov_eqn1}) to
linear order in $\phi-\phi_0$, and substitute in Eq. (\ref{gov_eqn2}). 
This allows us to find $\mu$,
\begin{eqnarray}
\mu=\frac{k_BT}{v_0}f_m'(\phi_0)-\frac14\Delta\rho\omega^2(R_2^2+R_1^2)
\end{eqnarray} 
where $f_m'=df_m/d\phi$. We notice that at the transition frequency and if
$\phi<\phi_c=1/2$, 
the largest
composition, found at $r=R_2$, crosses into the unstable part of the phase-diagram. This
means that $\omega_c$ occurs when $\phi=\phi_t$ and $r=R_2$ are substituted in Eq.
(\ref{gov_eqn1}). The transition composition $\phi_t$ at a given Flory parameter
$\chi$ (or temperature $T=B/\chi$) is given by
$N\chi=\ln\left(\phi_t/(1-\phi_t)\right)/(2\phi_t-1)$, and hence
we obtain the following expression for the critical demixing frequency 
\begin{eqnarray}\label{wc}
w_c^2=\frac{4k_BT}{v_0\Delta\rho}\frac{f_m''(\phi_0)}{R_2^2-R_1^2}(\phi_t-\phi_0)
\end{eqnarray} 
The notation $f_m''=d^2f_m/d\phi^2$ have been used. A similar expression exists for
$\omega_c^2$ as a function of $T-T_t$.

A plot of $\omega_c$ as a function of the distance from the transition (binodal) 
temperature $T_t$ is shown in Fig. 3. $\omega_c$ increases as the temperature is elevated
above $T_t$. If $\omega$ is larger than the plotted $\omega_c$, demixing 
occurs, otherwise the mixture displays only weak composition gradients. The numerical 
value of $\omega_c$ is quite low, and certainly achievable in conventional laboratory 
equipment -- $\omega_c=60$ rad/sec is equivalent to $572$ rpm.
\begin{figure}[h!]
\begin{center}
\includegraphics[scale=0.5,bb=15 180 545 585,clip]{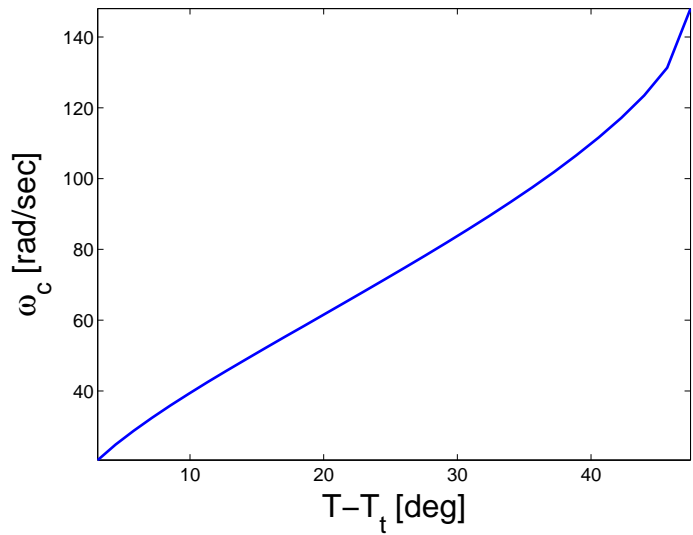}
\end{center}
\caption{Critical rotation frequency $\omega_c$ for demixing, as 
obtained from Eq. (\ref{wc}). $T$ is the temperature and $T_t$ is the transition 
(binodal) temperature for demixing without centrifugation. We took $\phi_0=0.2$, and 
all other parameters are as in Fig. 1.
}
\end{figure}

\section{Conclusion.} Centrifugation of mixtures, whether of simple liquids or polymers, 
is shown here to be an effective means to induce a phase-separation transition. 
As is quite intuitive, composition variations occur at any rotation frequency, 
because of the force which appears due to the different densities of the mixture's 
components \cite{sengers1,leung,moldover79,esterman79}.
The simple mean-field free-energy model we use reveals a surprising phenomenon: 
for temperature lower than the critical temperature $T_c$, there exists a critical
rotation frequency $\omega_c$. If the rotation is fast enough, $\omega>\omega_c$, a
demixing transition occurs where the high
density 
components is pulled toward the outer radius of the centrifuge, lower-density 
components are pulled in the other direction, and a sharp interface appears. 
At these high rotation frequencies, the 
centripetal force is large enough to cause large composition differences, which 
destabilize the homogeneous mixture in favor of a separated one. A somewhat similar 
phenomenon was predicted for liquid crystals under a gravitational field 
\cite{khokhlov}.

We note that demixing is also possible with spatially-varying electric fields.
The wedge capacitor, made up of two flat electrodes tilted
with respect to each other, is the analogue of the centrifuge. 
An electric field couples to the different permittivities of the liquid components 
$\Delta\eps=\eps_A-\eps_B$. The electrostatic energy density is
$\sim \Delta\eps E^2$ and in this geometry the
electric field varies like $r^{-1}$ \cite{TTL,TL_pre2005}. We thus find that
$\Delta\rho$ plays the role of the permittivity difference $\Delta\eps$, and $\omega r$
plays the role of an electric field. The critical voltage in the electrostatic case plays
the same role as the critical frequency $\omega_c$ in the present case.

We obtained numerical solution for a closed rotating container with a fixed 
composition, 
as well as an approximate analytical expression which compares well with the 
numerical results. The results we find are not specific to the 
mixing free-energy model utilized, Eq. (\ref{fm}); the transition appears in other 
models which
give a qualitatively correct phase-diagram in the absence of
centrifugation. In addition, the
$(\nabla\phi)^2$ term of Eq. (\ref{fm}) can be accounted for. It is expected to
decrease the composition gradients and may be important close to the critical point.
Numerical and analytical expressions for the case of an open 
system, having a connection to a liquid reservoir, and thus having an imposed
chemical potential, are easier to obtain and are also at hand. However, 
to experimental setup is not straightforward \cite{schaflinger}, and we chose to
leave it out.
%=================================================
% comment: consider adding the ``open system'' case in the second revision
% see apparatus in one of the papers; liquid comes from the bottom unrotating part
%=================================================

The demixing behavior described above is quite different from the one commonly 
observed in laboratory experiments, where the slow sedimentation of a substance
(e.g. polymers) in a solvent is simply accelerated by centrifugation. Our treaties show
that even {\it thermodynamically-stable} mixtures of simple and complex fluids can be
efficiently 
separated. It seems that this demixing transition can be used in many technological 
and chemical processes. For example, the reaction kinetics of two or more chemical  
species in a liquid environment can be controlled: if the mixture is homogeneous the 
reaction takes place everywhere, but when it is demixed the reaction can only take 
place at a thin boundary layer, and thus it has slower kinetics. 
Variations on this theme can also lead to accelerated kinetics: if one chemical 
species is an inhibitor for a reaction, after phase-separation it will 
have less contact with some of the liquid components, and the reaction can take 
place at a much higher rate.

Finally, we should stress that we used the simplest mean field model. This
approximation can be relevant for systems far from critical point and indeed we have shown
that even in such situations phase separation can be induced for rotation speeds easily
attainable by modern centrifuges. The kinetics of phase separation and of formation of
concentration profile within the sample seems to be an interesting open question that we
leave until some experiments are performed.

\section*{Acknowledgements}
We are indebted to Fran\c{c}ois Tournilhac for enthusiastic and stimulating
discussions. We thank Professor Jean-Marie Lehn for sharing with us his thoughts of
possible influence of gravitation forces on chemical equilibria and reactions.
This research was supported by the Israel Science foundation (ISF)
grant no. 284/05 and by the German-Israel Foundation (GIF) grant 
no. 2144-1636.10/2006.

\end{document}